\documentstyle{AGILEpsr}

\input epsf.sty
\input epsfig.sty
\input psfig.sty

\def \AAP #1 #2 {{\em Astron. Astrophys.\/} {\bf #1}, #2}
\def \AAL #1 #2 {{\em Astron. Astrophys. Lett.\/} {\bf #1}, L#2}
\def \AAR #1 #2 {{\em Astron. Astrophys. Rev.\/} {\bf #1}, #2}
\def \AAS #1 #2 {{\em Astron. Astrophys. Suppl. Ser.\/} {\bf #1}, #2}
\def \AJ #1 #2 {{\em Astron. J.\/} {\bf #1}, #2}
\def \ANNREV #1 #2 {{\em Ann. Rev. Astron. Astrophys.\/} {\bf #1}, #2}
\def \APJ #1 #2 {{\em Astrophys. J.\/} {\bf #1}, #2}
\def \APJL #1 #2 {{\em Astrophys. J. Lett.\/} {\bf #1}, L#2}
\def \APJS #1 #2 {{\em Astrophys. J. Suppl.\/} {\bf #1}, #2}
\def \APSS #1 #2 {{\em Astrophys. Space Sci.\/} {\bf #1}, #2}
\def \ASR #1 #2 {{\em Adv. Space Res.\/} {\bf #1}, #2}
\def \BAIC #1 #2 {{\em Bull. Astron. Inst. Czechosl.\/} {\bf #1}, #2}
\def \JSQRT #1 #2 {{\em J. Quant. Spectrosc. Radiat. Transfer\/} {\bf #1}, #2}
\def \MN #1 #2 {{\em Mon. Not. R. Astr. Soc.\/} {\bf #1}, #2}
\def \MEM #1 #2 {{\em Mem. R. Astr. Soc.\/} {\bf #1}, #2}
\def \PLR #1 #2 {{\em Phys. Lett. Rev.\/} {\bf #1}, #2}
\def \PASJ #1 #2 {{\em Publ. Astron. Soc. Japan\/} {\bf #1}, #2}
\def \PASP #1 #2 {{\em Publ. Astr. Soc. Pacific\/} {\bf #1}, #2}
\def \NAT #1 #2 {{\em Nature\/} {\bf #1}, #2}
\def \SAIT #1 #2 {{\em Mem.\ Soc.\ Astron.\ It.\/} {\bf #1}, #2}
\def \MESS #1 #2 {{\em The Messenger\/} {\bf #1}, #2}
\def \ASTRNACH #1 #2 {{\em Astron. Nach.\/} {\bf #1}, #2}
\def \AGPSR #1 #2 {{\em ASI Special Publication\/} {\bf #1}, #2}

\begin{opening}

\title{Low Energy X-ray Emission from Young Isolated Neutron Stars }
\author{M. Ruderman$^{1,2,3}$}
\institute{$^1$Department of Physics, Columbia University\\
$^2$Columbia Astrophysics Laboratory, Columbia University\\
$^3$Institute of Astronomy, Cambridge}
\date{} 
\end{opening}

\begin{document}

\def\lesssim{\mathrel{\hbox{\rlap{\hbox{\lower4pt\hbox{$\sim$}}}\hbox{$<$}}}}
\def\gtrsim{\mathrel{\hbox{\rlap{\hbox{\lower4pt\hbox{$\sim$}}}\hbox{$>$}}}}

\oddpagefooter{}{}{} 
\evenpagefooter{}{}{} 
\medskip  

\begin{abstract} 
A young neutron star with large spin-down power is expected to be closely
surrounded by an $e^\pm$ pair plasma maintained by the conversion of
$\gamma$-rays associated with the star's polar-cap and/or outer-gap
accelerators.  Cyclotron-resonance scattering by the $e^-$ and $e^+$
within several radii of such neutron stars prevents direct observations
of thermal X-rays from the stellar surface.  Estimates are presented for
the parameters of the Planck-like X-radiation which ultimately diffuses
out through this region.  Comparisons with observations, especially of
apparent blackbody emission areas as a function of neutron star age,
support the proposition that we are learning about a neutron star's
magnetosphere rather than about its surface from observations of young
neutron star thermal X-rays.

\end{abstract}

\medskip

\section{Introduction}

\def\lesssim{\mathrel{\hbox{\rlap{\hbox{\lower4pt\hbox{$\sim$}}}\hbox{$<$}}}}
\def\gtrsim{\mathrel{\hbox{\rlap{\hbox{\lower4pt\hbox{$\sim$}}}\hbox{$>$}}}}

A young, isolated, magnetized neutron star (NS) is expected to have
several potentially important sources of low energy ($\sim$keV) X-ray
emission (cf. Figure 1):

a) Thermal X-rays from the whole stellar surface as the NS cools down
after a very high temperature birth in a supernova explosion.

b) Thermal X-rays from its much smaller but hotter polar caps,
additionally heated by backflow to the stellar surface of extreme
relativistic electrons or positrons.  These leptons come from $e^\pm$
producing accelerators on the open $\vec B$-field-line bundles which
connect surface polar-caps to the spinning NS's distant ``light-cylinder."
These accelerators may be just above the NS surface (``polar-cap
accelerators") or in the NS's outer-magnetosphere (``outer-gap
accelerators").

c) Synchrotron radiation from newly created $e^\pm$ pairs as the $e^-$
and $e^+$ radiate away their initial momenta perpendicular to local $\vec B$.

The angular and spectral resolution achieved by the {\it Chandra} and
{\it XMM} satellites allows discrimination of such NS radiation from that
of the supernova remnant in which the youngest (age $t \lesssim 10^4$yrs)
NSs are usually embedded, and also thermal components (a and b) from the
power low component (c); cf. Figure 2. In Section 2 we consider rather
dramatic differences between observed thermal X-ray emission from young
isolated hot neutron stars and the expected emission if the NS (near)
magnetospheres were essentially empty. Section 3 discusses consequences
for this emission if the NS magnetosphere within several NS radii of the
stellar surface has an $e^\pm$ pair density $(n_\pm)$ large enough to
give multiple cyclotron-resonance scattering of NS thermal X-rays before
they ultimately escape or scatter back to the star.  Section 4 estimates
the steady state $n_\pm$ and its consequences for observed NS thermal
spectra.  Section 5 summarizes some conclusions and further problems.

\begin{figure}
\epsfysize=6cm 
\centerline{\epsfbox{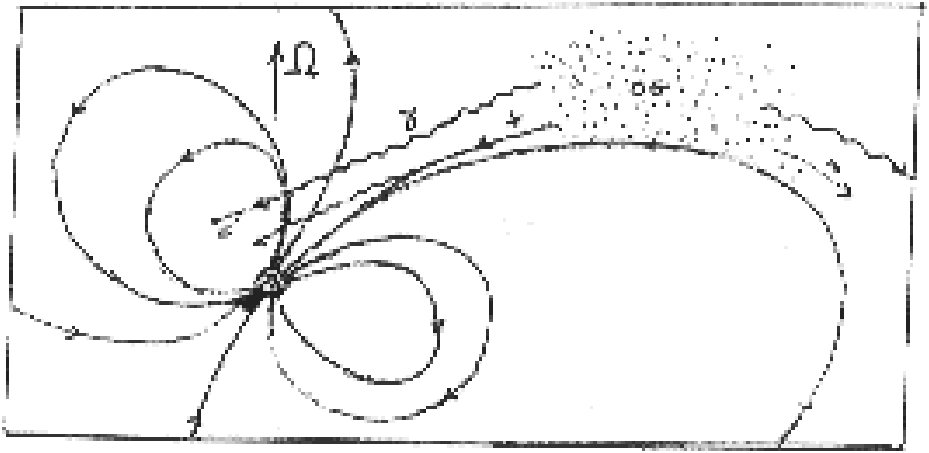}}
\caption[h]{An outer-gap (OG) accelerator, extending along the open field line bundle on
both sides of its intersection with the null-surface where $\Omega \cdot {\vec B} = 0$.
GeV $\gamma$-rays directly from this accelerator may produce $e^\pm$ pairs close to the
neutron star. $10^2$ MeV $\gamma$-rays, curvature radiated by TeV $e^+(e^-)$ flowing out
of this accelerator down to the star, will be a very strong source for such $e^\pm$.  On the
other side of the star a polar cap (pc) accelerator may also be a strong source for $\gamma
\rightarrow e^+ + e^-$. (Because $e^\pm$ pairs from OG accelerators can quench PC
accelerators and vice versa, it is unlikely that a particular bundle of open field-lines
would sustain both.)}
\end{figure}

\begin{figure}
\epsfysize=6cm
\centerline{\epsfbox{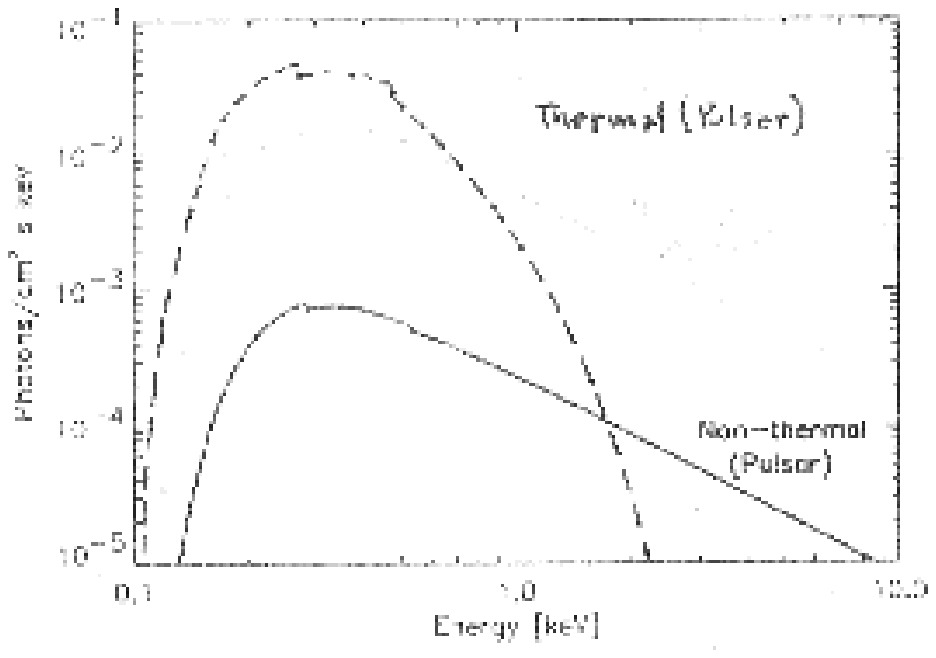}}
\caption[h]{X-ray count flux from the Vela pulsar (after removal of nebular emission).
(After Mori et al. 2002.)}
\end{figure}

\section{Observations and models}

Reported thermal X-ray luminosities $(L^T_X)$ of young isolated neutron
stars are shown as a function of NS ``age" ($t$) in Figure 3.
(The ``age'' is generally that  of the SNE envelope around the NS when
that can be estimated.  For some of the youngest NSs, J2323, J0852, J1617, and J0821
spin-period $P$ with spin-down-rate $\dot P$ are not known so the canonical spin-down age
$(P/2\dot P)$ cannot be used.)  The computed thermal $L^T_X$  is that for the simplest,
initially very hot, $1.4 M_\odot$ neutron star model: a core consisting solely of neutrons,
protons and extreme relativistic electrons, surrounded by a relatively
thin more poorly conducting crust which keeps the NS surface temperature
($T$) very much less than that at the crust-core boundary
$(T_c)$.  
Although model $T$ depends upon assumed NS surface detail, $L_X^T$ from core-cooling should
not as long as $T_c\gg T$.
Disagreement between model estimates and observations is not great, and usually compatible
with uncertainties in the distance to the NS.  (PSR 1055, for example, may well be half as
far away as the value used in Figure 3, decreasing its $L_X^T$ by a factor four.  Some of
the extremely young NSs may have apparent $L_X^T$ which are strongly diminished when they
are observed along particular lines-of-sight, as discussed in Section 5e.)  Comparison
between predictions and observations of Figure 3 do not discourage attribution of those
$L_X^T$ to the expected NS cooling.

However, there are two great surprises in the observed thermal
spectra of the NSs.

1) None of the spectral features calculated for NS surface compositions
other than H (and, presumably, He) are observed (with the remarkable
exception of 1E1207 which may have a very different explanation to be
discussed elsewhere).  Indeed, a blackbody spectrum is generally an
acceptable fit to observations and radiation parameters (surface
temperature and radiating area) have usually been calculated for it.

\begin{figure}
\epsfxsize=8cm
\centerline{\epsfbox{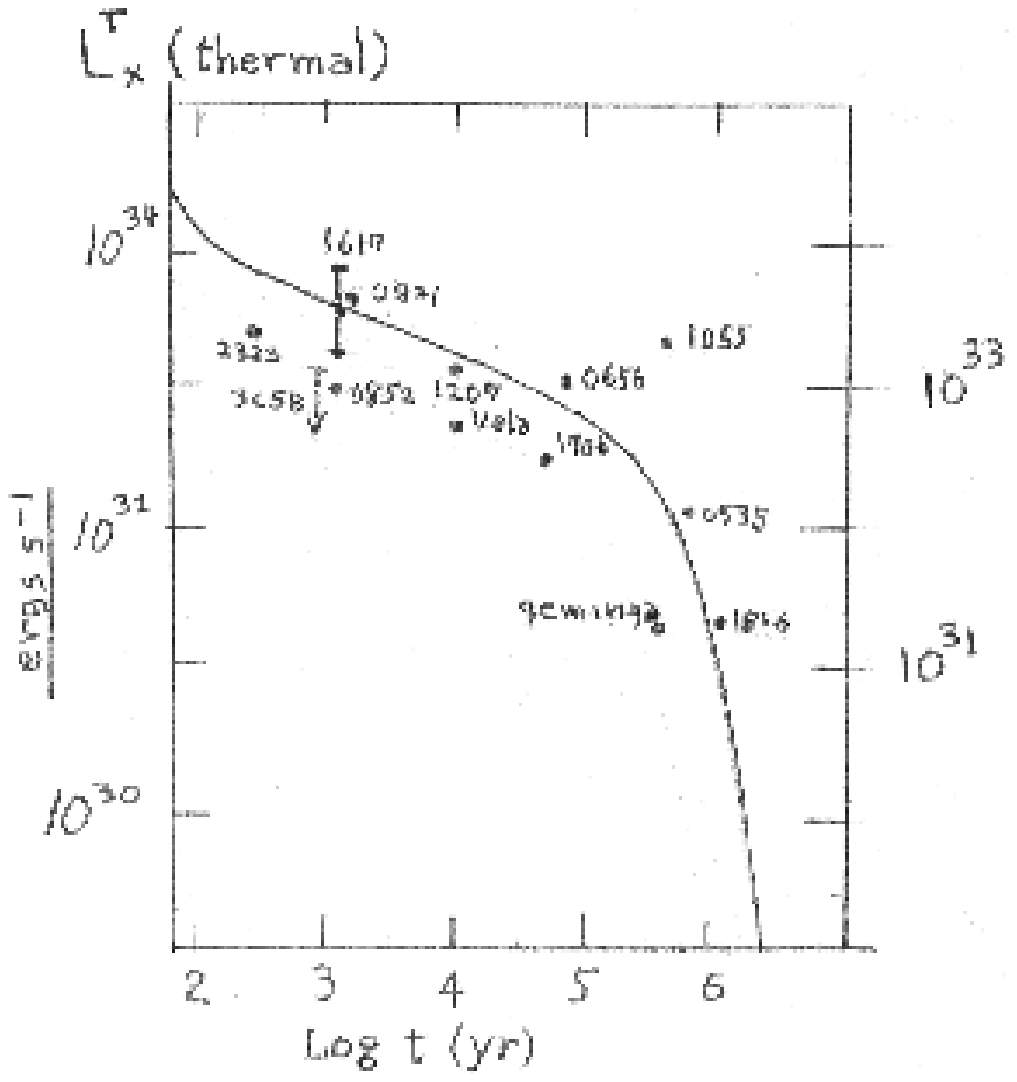}}
\caption[h]{Thermal X-ray luminosity $(L_X^T)$ of young isolated canonically magnetized
neutron stars as a function of stellar age $(t)$ Observed $L_X^T$ are from summaries of
{\it XMM} and {\it Chandra} results (Becker \& Aschenbach 2002; Pavlov et al. 2002)
supplemented with data from Gotthelf et al. (2002) for PSR B1706, McGowan et al. (2003) for
J0538, Mori et al. (2002) for Vela, and Wang \& Halpern (1997) for Geminga.  The solid line
model curve for $L_X^T$ to which these data are compared is based upon Page (1998); cf.
Tsuruta (1997).}
\end{figure}

2) The effective blackbody radiating area $(A_{BB})$ has the magnitude
and age dependence indicated in Figure 4, hugely different from a simple
age independent $A_{BB} \sim A_{NS} \equiv 4\pi (R_{NS})^2 \sim 4\pi
(10~{\rm km})^2 \sim 10^{13}$cm.  Instead, observations give
$$ {A_{BB}\over A_{NS}} \sim 3\cdot 10^{-2} \left( {t\over 10^4{\rm
yrs}}\right)\eqno(1)$$
for $t \lesssim  10^5$ yrs; and $A_{BB}/A_{NS} \sim 0(1)$ only for older
NSs.  If, despite the support of Figure 3 for the observed $L_X^T$ being
mainly powered by core cooling, it is assumed that the dominant sources of
$L_X^T$ are the hotter polar caps, the difference between expectations
and observed black body areas becomes even greater:  NSs with central
dipoles (or uniform interior magnetization) have polar caps of area
$(A_{pc})$ with 
$${A_{pc}\over A_{NS}} \sim {5\cdot 10^{-5}\over P{(\rm sec)}}.\eqno(2)$$ 
This ratio is shown as the downward sloping dashed line of Figure 4,
differing even more dramatically in both slope and magnitude from
observations than does $A_{BB}/A_{NS}\sim 1$.

\begin{figure}
\epsfysize=10cm
\centerline{\epsfbox{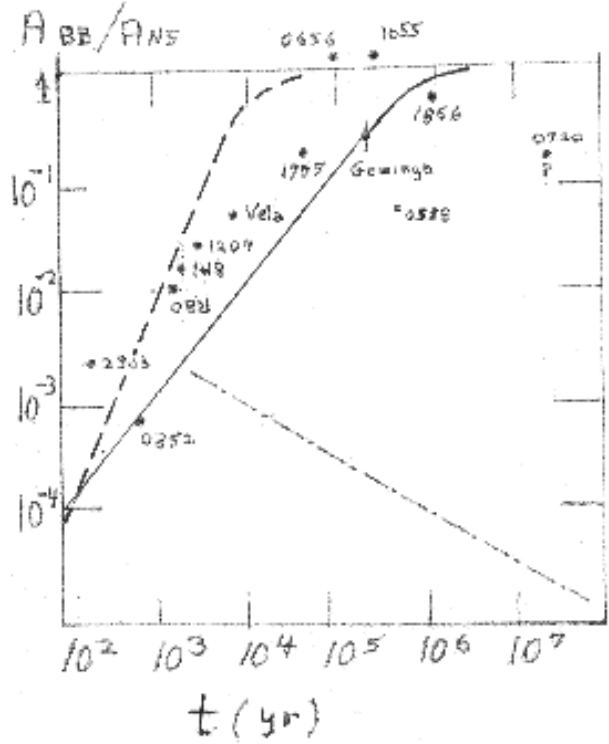}}
\caption[h]{Blackbody fits for thermal radiation areas of isolated neutron stars $(A_{\rm
BB})$, in units of a nominal neutron star surface area $A_{NS} = 4\pi (10{\rm km})^2$, as
a function of neutron star age. (References for these data are the same as those for Figure
3.) The decreasing dash-dot diagonal is the model area of a hot polar cap of a central
dipole.  The upper dashed line is from the $\tau = 10^{-3}$s $e^\pm$ outflow model of
Section 4; the lower solid line is from the $\tau = 0$~s model in
which 
$e^\pm$ flow before $e^+ + e^- \rightarrow \gamma + \gamma$ annihilation is assumed to
be ignorable.}
\end{figure}

If it is assumed that the NS surface, instead being a perfect absorber,
has an H-atmosphere, $L^T_X$ is not much changed.  The associated
affective radiating areas for near keV X-rays still grow almost linearly
with $t$ but now would have very significantly greater magnitudes. 
Typically the two different inferred NS surface temperatures, $T_{BB}$
and $T_H$, differ by about a factor two; $T_{BB} = (1.5 - 3)T_H$ (Pavlov
et al. 2002; Becker \& Aschenbach 2002) and have area ratios inversely
proportional to the surface $T^4$.  Then for putative H-atmospheres when
only blackbody parameters have so far been reported it seems plausible
to estimate $A_H \sim 2^4A_{BB}$: the youngest NSs would still have
$A_H/A_{NS}\ll 1$.  The dramatic rise and subsequent leveling off in
radiating areas would still remain and require explanation.


\section{Thermal X-rays observed from a hot strongly magnetized neutron star
embedded in a dense $e^\pm$ plasma}

A surrounding $e^\pm$ plasma with number density $n_\pm = n_+ = n_- <
10^{17}$cm$^{-3}$, extending out to several stellar radii around a NS,
would not give important Thomson cross section scattering of thermal
X-rays emitted from the NS surface.  However, because the thermal X-ray's
energy ($E_X =\hbar \omega \sim$ keV) is less than the $e^-$ and $e^+$
cyclotron-resonance energy at the stellar surface ($\hbar \omega_B =
eB\hbar/mc
\sim 10 B_{12}$keV), an emitted thermal X-ray will pass through a
cyclotron-resonance where $E_X - \hbar \omega_B(r)$ within a few stellar
radii of the emitting surface (Wang et al. 1998, Zhu \& Ruderman 1997,Dermer \& Sterner
1991, Mitrofanov \& Pavlov 1982).  The resonant polarization-averaged X-ray
scattering cross section there is well approximated as 
$$\sigma_\pm \sim
{4\pi^2e^2\over mc} \delta [\omega - \omega_B(r)].\eqno(3)$$
We assume below that equation (3) also adequately approximateds the
scattering cross section for any polarized x-ray if the averaging is over
its $e^-$ and $e^+$ scattering cross sections. (This fails when the
photon's momentum is almost perpendicular to local $\vec B$ together with
its electric polarization being almost parallel to $\vec B$.  However,
when the optical depth for such resonance scattering is very large, as in
the main applications discussed below, or the near-star $\vec B$-field is
very irregular, the limit to accuracy from the use of Equation 3 does not
substantively alter the conclusions which will be based upon it.)

With the $\sigma$ of Equation 3, the optical depth for
cyclotron-resonance scattering $(OD)$ of an energy $E_X$ X-ray is about 
$(\lambda_X/ {e^2\over mc^2}) \sim 10^5$ times that for non-resonant
Thomson scattering:
$$ OD = \int^\infty_{R_{NS}} \sigma_\pm n_\pm dr \sim { 4\pi^2 e^2 r_B \hbar 
n_\pm\over 3 mcE_X} \eqno(4)$$
as long as $r_B$, the distance from the $NS$ center where $\hbar \omega_B
= E_X$, satisfies $r_B > R_{NS}$.  (The denominator factor 3 is based on
$B$-field $r$-dependence of a central dipole.)

The cyclotron-resonance backscatter radius $r_B \sim 3 R_{NS}$ for an
$E_X \sim 1$keV X-ray emitted from a NS with surface dipole field $3\cdot
10^{12}$G.  This backscattering region around $r\sim 3R_{NS}$ forms a
reflecting, slightly leaky {\it Hohlraum} container holding in radiation
from the NS.  Its X-ray transmission coefficient
$$\hat t \sim \left( {2\over 2 + OD}\right),\eqno(5)$$
and the container X-ray reflection coefficient $\hat r = 1 - \hat t \sim
OD (OD+2)^{-1}$.
When $\hat t \ll R^2_{NS}/r^2_B \sim 10^{-1}$, an X-ray emitted from the
NS surface will almost always be reflected by the {\it Hohlraum}
container and then reabsorbed by the NS before escaping the 
{\it Hohlraum}.  In this regime the {\it Hohlraum}-contained X-rays
acquire a Planck-blackbody energy spectrum independent of the NS
surface's emission and reflection properties.  It is sensitive only to
the (now higher) surface temperature of the NS.

In a steady state the total outflow luminosity is unchanged by the $e^\pm$ plasma in which
the NS is embedded.  Equivalently, the container has a surface emissivity $\hat e \approx
\hat t = 1 - \hat r$ and
$$L_X \equiv A_{BB}\sigma_{SB}T^4_{BB} = 4\pi r^2_B \hat e \sigma_{SB} T_{BB}^4\eqno(6)$$
(with $\sigma_{SB}$ the Stefan Boltzmann constant).

Then from Equations (5) and (6)
$$A_{BB}\simeq \hat e 4\pi r^2_B \simeq {8\pi r^2_B\over OD}\eqno(7)$$
as long as the computed $R_{NS}$ of Equation (7) gives an $A_{BB} \ll 4\pi R^2_{NS}$.
In the opposite limit, as $OD$ drops below $(r_B/R_{NS})^2/2$, $A_{BB}\rightarrow 4\pi
R^2_{NS}$.  If $A_{BB}$ varies greatly with $E_X$ observed emission out from the {\it
Hohlraum's} container would have a spectral distribution significantly different from the
interior Planck one.

\section{Inferred radiation areas ($A_{BB}$) and neutron star spin-down age}

In a steady state the production rate of $e^\pm$ pairs near a spinning-down NS $(\dot
N_\pm)$ needed to maintain a constant local pair density $n_\pm$ in the relevant
cyclotron-resonance scattering region is balanced by the sum of two local pair removal
rates:

a) Local annihilation into $\gamma$-rays $(\dot n^a_\pm)$ with
$$n_\pm^a \sim 2\pi \left( {e^2\over mc^2}\right)^2 c \left( n_\pm \right)^2;
\eqno(8)$$
(assuming nonrelativistic $e^+ - e^-$ relative velocities along $\vec B$ and anti-parallel
magnetic moments along the strong local $\vec B$ have been achieved).

b) Pair outflow $(\dot n^f_\pm)$ from strong push of incident thermal X-rays on the
continually replenished cyclotron-resonant container of the {\it Hohlraum}
give a finite residence time $(\tau)$ there:
$$ \dot n^f_\pm \sim - {n_\pm\over \tau}.\eqno(9)$$
One estimate for the $\tau$ follows from the assumption that very quickly
after the pair creation  the $e^\pm$ speed along local $\vec B$ becomes
such that the push on $e^-  + e^+$ is perpendicular to local $\vec B$ 
(about $10^{10}{\rm cm\ s}^{-1}$). Then to move along that $\vec B$ through
the container thickness $(\sim r_B \sim 3R_{NS} \sim 3 \cdot 10^6{\rm cm}$)
 takes
$$\tau \sim 10^{-3}{\rm s}.\eqno(10a)$$
[Pair inflow from NS gravitational pull could be important if that pull were greater than
the outward push by contained {\it Hohlraum} X-rays.  In the models presented here the
pressure for that outward push is about $(OD/2)(L_X^T/4\pi r^2_Bc)$.  The oppositely
directed pull from the weight of the $n_\pm$ around $r_B$ which constitute the
container (Equation 4) is only $10^{-3} (10^{33}{\rm erg\ s}^{-1}/L_X^T)$ as large. 
Therefore we do not consider it further here.  This interpretation does not support the
gravity dependent $e^\pm$-layer structure and X-ray spectra proposed for the Crab pulsar
by Zhu and Ruderman (1997); cf. also Wang et al. (1998).]

At the other extreme, if this local $\vec B$ near the NS is still so ``complicated'' that
$e^\pm$ is kept from flowing out before annihilation,
$$\tau \sim \infty.\eqno(10b)$$

The near-star pair input rate may be even more difficult to estimate than the annihilation
rate.  Pairs on closed field lines very near a young NS $(r \lesssim 3 R_{NS})$ can be
created by  a) MeV - GeV $\gamma$-rays which enter into that region from outer-gap
accelerators; b) MeV - $10^2$MeV curvature $\gamma$-rays from TeV $e^+(e^-)$ flowing down
to a NS polar cap on $\vec B$-field lines connecting the polar cap to the outer-gap
accelerators; and c) 1 - $10^2$MeV $\gamma$-rays generated by polar cap accelerators on
open-field lines, and then bent over to cross near-surface closed-field lines by the strong
NS gravitational pull on them (a source which becomes large and important when the usual
central dipole model for $\vec B$ is replaced by a more realistic one without axial
symmetry).

A further complication in calculating $\dot N_\pm$ near the NS is a possible second
generation of $e^\pm$ production by GeV $\gamma$-rays from outer-gap accelerators.  The
$e^\pm$ pairs they create, if  these $\gamma$-rays pass are within $10 R_{NS}$ of the NS,
can themselves be the source of energetic ($E_\gamma > 1$MeV) synchrotron $\gamma$-rays. 
These, in turn, would create more pairs if they pass into the much stronger $\vec B$ much
closer to the star (Cheng \& Zhang 1999; Chang et al. 1998; Yakovlev et al. 2002).  A rough
estimate for the Vela pulsar suggests that $\dot N_\pm \sim 10^{37} - 10^{38}$s$^{-1}$ for
this pulsar.

A more empirical, and probably more reliable, estimate of $\dot N_\pm$ near the NS comes
from identifying the Vela power law component of Figure 1 with the x-ray synchrotron
radiation from the immediate quenching of $e^-$ and $e^+$ perpendicular momenta of all
newly created $e^\pm$ pairs in the strong local $\vec B$ (Wang et al. 1998). The power law
synchrotron luminosity from the local field where the $e^\pm$ are created by $\sim 10^2$MeV
$\gamma$-rays has a total power law luminosity $(L_X^{PL})$ for 2 keV $< E_X < 10$ keV
$$L_x^{PL} (2 - 10 {\rm keV}) \sim \dot N_\pm mc^2\left( {8 {\rm keV}\over \hbar\omega_B}
\right)^{1/2} \sim 10^{-5} \dot N_\pm {\rm erg\ s}^{-1},\eqno(11)$$
where the local $\vec B$ field is assumed to be around $10^{10}$G so that $\hbar
\omega_B\sim 10^{-1}$keV.
If the $L_X^{PL}$ is indeed from such synchrotron  radiation, its X-ray photon flux
spectrum between 2 and 10 keV should be $\sim E_X^{-\Gamma}$ with $\Gamma \sim 1.5$
as long as all relevant $\hbar \omega_B \lesssim 2$ keV and maximum synchrotron radiation
energy $E_X > 10$ keV. [Although reported $\Gamma$ ranges do not always include this value
it is compatible with many
examples: 1.25 -- 1.45 for Vela, 1.4 -- 1.5 for PSR 0656, 1.6 for the Crab (Becker \&
Aschenbach 2002) and 1.66 for PSR 1055 (Pavlov et al. (2002).]

Observed (2 -- 10 keV) $L_X$ as a function of  NS ages are shown in Figure 5. (In
this range for $E_X$, $L_X$ is  expected to be dominated by its power law component.)

\begin{figure}
\epsfysize=6cm
\centerline{\epsfbox{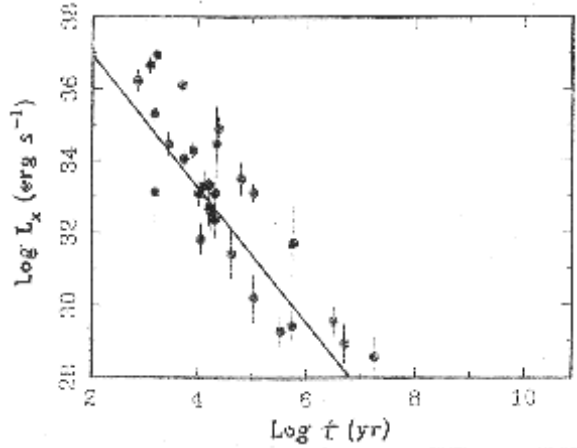}}
\caption[h]{X-ray luminosity $(L_X)$ 
for 2 keV $< E_X < 0$ keV
of canonically magnetized isolated neutron stars as a
function of their ages (Possenti et al. 2002).  The solid curve is Equation 12:
$L_C = 10^{33}(10^4 {\rm yrs}/t)^2{\rm erg\ s}^{-1}$.}
\end{figure}

We approximate a fit to these data by the line shown there:
$$L_X^{PL} \sim L_X = 10^{33} \left( {10^4{\rm yrs}\over t}\right)^2{\rm erg\
s}^{-1}.\eqno(12)$$
When combined it with the defining relations $B^2_d \equiv R^{-6}c^3\Omega^{-4} \times$
(spin-down power $| I\Omega \dot\Omega |$) and NS spin-down age $t=-\Omega/2\dot\Omega$,
Equation 12 becomes $L_X \sim 4 B_d^2 R^6_{NS} I^{-2} \times |$ spin-down power$|$.
For canonical pulsars with $B_d \sim 10^{12}$G it is then roughly equivalent to the
empirical Becker-Tr\"umper relation (Becker \& Tr\"umper 1997; Seward \& Wang 1998;
Possenti et al. 2002.)
$L_X \sim 10^{-3}|{\rm {\hbox{spin-down}}\ power}|$.

To calculate the needed $n_\pm$ by combining the total pair creation rate $\dot N_\pm$ from
Equations 11 and 12 with the local disappearance rates of Equation (8, 9 and 10), it is
necessary to estimate the volume $(V_\pm)$ in which the pair creation
rate of $\dot N_\pm$ is
achieved.  This depends upon the characteristic $\vec B$ in which the $\gamma + \vec B
\rightarrow e^\pm + \vec B$ occurs, the same $\vec B$ as that of $\omega_B$ in Equation 11. 
For that
$\hbar
\omega_B \sim 10^{-1}$keV, corresponding to $e^\pm$ production beginning at $B\sim
10^{10}$G by characteristic $10^2$MeV  curvature $\gamma$-rays from the TeV $e^-(e^+)$
flow outward from a polar-cap accelerator and inward from an outer-gap one,
$$V_\pm (\propto \omega_B^{-3}) \sim 10^{21}{\rm cm}^3.\eqno(13)$$
Combining Equations 7 and 4 with $r_B = 3\cdot 10^6$cm and $E_X =1$keV, together with
Equations 8, 9, 11, 12, and 13, gives $n_\pm$, $OD$, and the two $A_{BB}/A_{NS}$ model
curves shown in Figure 4.  To obtain the solid curve we pretend that equation 10b is the
appropriate entry for Equation 9, i.e., a very complex closed B-field structure to $r\sim
r_B$.  The opposite $\tau = 10^{-3}$s approximation of Equation 10a gives the upper dotted
line.  The truth may well lie somewhere between the curves shown.

\section{Some conclusions, speculations, and problems}

The model results for estimated $A_{BB}$ magnitudes and their age dependence are
qualitatively similar to the observed ones of Figure 4 over several orders of magnitude. 
This gives considerable support to our central proposition. Strongly magnetized NSSs with
large spin-down powers continually produce enough $e^\pm$ pairs on the closed B-field lines
in their strongly magnetized near-magnetospheres that X-ray cyclotron-resonant scattering
there preempts direct observation of the NS surface.

Although comparisons between these over-simplified models and NS thermal X-ray observations
may not be quantitatively reliable they encourage further model-based interpretations of
existing thermal X-ray observations.

If the $\tau = 10^{-3}$s model (uppermost curve) in Figure 4 leads to a reasonable
approximation for $e^\pm$ cyclotron-resonance opacity around a young neutron star, that
opacity alone would not explain why the NSs older than $10^4$ yrs wold have $L_X^t$ fits to
near blackbody spectra and $A_{BB}/A_{NS}$ so much less than unity.  These two features
suggest that for those NSs (but not for younger ones) it may be more appropriate to fit
observed thermal spectra with unobstructed emission from H or He atmospheres. (This seems,
however, often to give optimal fits with $A_{H}/A_{NS}$ enough larger than unity to
introduce new problems.)

If the $\tau \rightarrow \infty$ model is a better description, surrounding $e^\pm$ plasma
blackbody-like emission from a slightly leaky $e^\pm$ {\it Hohlraum} container would remain
appropriate for very considerably older NSs (lower curves of Figure 4).  However, at least
one of these, J0538, would be expected to have such small surrounding $e^\pm$  opacity
that the observations still suggest its interpretation as a NS emitting unobstructed thermal
X-rays from an H (or He)-atmosphere with
$$ {A_H\over A_{NS}} \sim \left( {T_{BB}\over T_H}\right)^4 { A_{BB} \over A_{NS}}
\sim 2^4 { A_{BB}\over A_{NS}} \sim 1.\eqno(14)$$

A more accurate description of the large $n_\pm$ around very young magnetized spinning-down
NSs will be needed for many observational details.

a) Variation of $n_\pm$ and thus $OD$, with (angular) location.  This would give spin-phase
modulation to the observed thermal flux.

b) The dependence of $r_B^2/OD$ on X-ray frequency, causing some departure from a Planck
spectrum in emission. (If the $\dot N_\pm$ which sustains $n_\pm$ is inside $r_B$ and
$n_\pm$ outflow through relevant $r_B$ conserves $r\pi r^2n_\pm$. Equation 4 does give
$r\pi r^2_B/OD$ independent of 
X-ray energy when $B\propto r^{-3}$.) Another neglected departure of thermal X-ray emission
from exact Planck spectra comes from significant $e^+/e^-$ flow velocities along $\vec B$:
$e^+/e^-$ recoil is very small in each cyclotron-resonance X-ray scatter $(E_X/mc^2 \ll 1)$
but not X-ray energy boost or degradation.

c) Inclusion of backscatter of hotter polar-cap X-rays which suppresses discrimination
between the separate contributions to $L_X$ of polar cap emission and general surface
cooling.

d) Correlations among $L_X^{PL}$, $A_{BB}$, and dipole moment for individual NSs.  For
example, special consideration must be given to relating $\dot N_\pm$ to $L_X^{PL}$ from
pulsars with exceptionally large $B(\sim 10^6{\rm G})$ near their light cylinders.  When
such pulsars have strong outer-gap accelerators in that region, GeV (curvature)
$\gamma$-rays + keV X-rays produce pairs there with very strong initial synchrotron
radiation in the 2 -- 10 keV range.  For these $e^\pm$, local $\hbar \omega_B \sim
10^{-2}$eV so Equation 11 becomes $\dot N_\pm \sim 10^3L_X^{PL}$ (erg s$^{-1}$) instead of
the much larger inferred $\dot N_\pm$ from $L_X^{PL}$ if the $e^\pm$ pairs
are produced very near the NS.  This distinction should be important for the Crab pulsar
with its exceptionally large light-cylinder B ($\sim 10^6$G).  The Crab pulsar's observed 
$L_X^{PL}$ (2 -- 10 keV), $\sim 10^{36}$erg~s$^{-1}$, should and does lie well above the
$L\propto t^{-2}$ line in Figure 5.  Its $L_X^{PL}$ should not be used without adjustment to
predict $\dot N_\pm$ in the Crab pulsar magnetosphere and especially not in the
inner-magnetosphere near the star.  (A total crab pulsar $\dot N_\pm$ inferred from
Equation 11, without taking into account its large expected outer-magnetosphere contribution
to $L_X^{PL}$, would exceed observational upper limits for the Crab pulsar's half MeV
annihilation $\gamma$-rays from $e^+ + e^- \rightarrow \gamma + 
\gamma$, (0.5 -- 1) $\times 10^{40}$s$^{-1}$ (Ulmer et al. 2001; Massaro et al. 1991; Zhu \&
Ruderman 1997).

e) Differences between total $L_X^T$ and that inferred from observations because the spin
cycle averaged $OD$ may vary greatly for observers along different lines-of-sight to the NS
center.  In particular, there is the possible escape of {\it Hohlraum} X-rays through an
open field-line bundle ``hole'' (area $\sim \pi r_B^3\Omega c^{-1}$) in the {\it Hohlraum}
container.  (Because of the large very relativistic particle flow in the bundle sustained
by the accelerator somewhere on it, $OD$ along it can be greatly suppressed at $r\sim r_B$
for thermal x-ray frequencies.  Then, for pulsars with relatively large ``hole" areas and
large
$n_\pm$ elsewhere, {\it Hohlraum} X-rays may mainly pass out through the container's
polar-cap-like ``holes'' rather than diffuse out through the container's reflecting ``wall".
When this obtains, the concentration of $L_X^T$ outflow through such holes could increase
its apparent value by about a factor 2, or decrease  it by a very much greater factor,
depending on polar cap locations and that of the line-of-sight to the observer. (The
rapidly spinning NS in 3C58 (J0205 with $P=66$ ms) with very large spin-down power
$(2\times 10^{37}{\rm erg\ s}^{-1})$ would be expected to have a very large $\dot N_\pm$ and
thus $n_\pm$.  X-ray outflow through container holes should be greater than the diffusive
one.  To an observer for whom the container holes are largely hidden during the full NS
spin cycle  most of that NS's $L_X^T$ could be directed away from the observer.  If so,
the inferred $L_X^T$ magnitude would be greatly reduced from its true value.  This may well
account for not yet observing any $L_X^T$ from it (cf. Slane et al. 2002).

Effects of $\dot N_\pm$ may be complicated and difficult to evaluate quantitatively but
their consequences may be crucial for interpreting observations of low energy X-ray fluxes
from young isolated neutron stars.

\acknowledgements

I am pleased to thank many colleagues for patient listening and informing, especially W.
Becker, E. Gotthelf, C. Hailey, J. Halpern, D. Helfand, C. Ho, K. Mori, F. Paerels, G.
Pavlov, M. Rees,  J. Tr\"umper, and M. Weisskopf.

\end{document}